%% file: main.tex
\documentclass[twocolumn,floatfix,showpacs,preprintnumbers,amsmath,amssymb,showkeys,prb]{revtex4-1} 

\usepackage[T1]{fontenc}
\usepackage[english]{babel}
\usepackage{graphicx}
\usepackage{amsmath,amssymb,mathrsfs}
\usepackage{hyperref}
\usepackage{caption}
\usepackage{mypackages}
\usepackage{mymacros}
\usepackage{todonotes}
\usetikzlibrary{matrix}

\begin{document}

\widetext

\title{Optical spectra of rare-earth nickelates}
\input author_list.tex

\date{\today}

\begin{abstract} 
Based on the electronic band structure obtained from first principles DFT calculations, the optical spectra of yttrium and neodymium nickelates are computed. We show that the results are in fair agreement with available experimental data.  We clarify the electronic transitions at the origin of the first two peaks, highlighting the important role of transitions from $t_{2g}$ states neglected in previous models. We discuss the evolution of the optical spectra from small to large rare-earth cations and relate the changes to the electronic band structure.
\end{abstract}

\maketitle

\section{\label{sec:level1}Introduction}
$ABO_3$ perovskites form a broad and technologically important family of compounds\cite{Khomskii2014,Imada1998,Zubko2011,Yuan2018}. Depending of A and B cations, they can exhibit a wide range of distinct behaviors -- ferroelectricity, piezoelectricity, multiferroism, metal-insulator transition, ... -- making them attractive for various device applications \cite{Grisolia2016,Hwang2012}. 

Amongst perovskites, rare-earth nickelates, RNiO$_3$, show on cooling an unusual and very complex phase diagram combining structural, electronic and magnetic phase transitions.  Except for LaNiO$_3$ which remains always in a metallic paramagnetic rhombohedral phase \cite{Scherwitzl2011}, other RNiO$_3$ compounds  typically show on cooling concurrent structural and electronic transitions from an orthorhombic $Pbnm$ metallic phase to a monoclinic $P2_1/n$ insulating phase \cite{Torrance1992,Medarde1997,Catalan2008}. The structural distortion accompanying this metal-insulator transition (MIT) consists in  a {\it breathing} distortion of the corner-shared NiO$_6$ oxygen octahedra, yielding a 3-dimensional checkerboard arrangement of large ($Ni_L$) and small ($Ni_S$) cages compatible with a related charge ordering.  The temperature of the MIT is moreover directly linked to the $R$ cation size, or equivalently to the Goldschmidt tolerance factor\cite{Goldschmidt1926}, and ranges from 100 K for PrNiO$_3$ to 600 K for YNiO$_3$ \cite{Barman1994,Imada1998}. Additionally, these compounds also show a low-temperature magnetic transition from paramagnetic to an E'-type antiferromagnetic order \cite{Garcia-Munoz1992}. Depending again of the cation size, this magnetic transition can be either independent (small cations) or concurrent (large cations) to the structural and electronic transitions.  

There has been a long debate about the origin and mechanism of the MIT in these compounds\cite{Catalano2018}. It was sometimes considered as a Mott-Hubbard transition\cite{Mizokawa2000}. In line with the appearance of the breathing lattice distortion, it was also related to a formal $d^8/d^6$ charge ordering at the nickel sites, requiring a negative U \cite{Mazin2007,Seth2017}. It was however further shown that the electronic configuration is better described by a  $d^8/d^8L^2$ occupation for the two nickels. From this, the MIT has been seen as a site-selective Mott transition \cite{Park2012,Subedi2015}. Relying on its strong dependence with oxygen rotation motions, the MIT was also recently re-interpreted as a structurally triggered Peierls transition, linked to an electron-phonon coupling \cite{Mercy2017}. Although some controversies remain, these views are not necessarily in contradiction and it seems now accepted that a proper picture\cite{Catalano2018} should consider together electronic and atomic degrees of freedom \cite{Mercy2017,Peil2019,Hampel2019}.

One concrete way of probing experimentally the metal-insulator transition of nickelates is from optical conductivity measurements \cite{Ruppen2015,Stewart2011,Toriss2017}. Indeed, the opening of a gap and the changes in the electronic structure of the conduction bands taking place at the MIT significantly affect the optical spectrum. Moreover, the effect of magnetism on the optical conductivity can also be probed \cite{Georgescu2019}. One limitation of optical conductivity measurements is however that, although they successfully provide a global view on the electronic structure and its modifications, they do not provide direct access to a "state-by-state" analysis. Proper interpretation of optical spectra therefore typically requires to combine measurements with electronic band structure calculations allowing to assign measured peaks to specific electronic transitions. 

Optical spectra of nickelates \cite{Ruppen2015} have, up to now, only been reported on thin films\cite{Ruppen2015,Stewart2011,Toriss2017}. They appear characterised, close to the conductivity edge, by a first main peak with a shoulder followed by a second peak in a range of 2 eV. Other more complicated features then appear at higher energies, but are however sensitive to the substrate \cite{Toriss2017}. Relying on DMFT calculations\cite{Ruppen2017} restricted to Ni-$e_g$ states only, the first peak was assigned to inter-site electronic transitions from $Ni_L$ to $Ni_S$ sites at the Mott gap while the second peak was interpreted as coming from the opening of a Peierls pseudogap at larger energy linked to the breathing distortion\cite{Ruppen2015,Subedi2015}. 

As recently discussed, density functional calculations reproduce rather the MIT transition as a Peierls-type transition. Relying on that, it was recently stated \cite{Peil2019} that the DFT band structure is not compatible with the observed optical spectra of nickelates, further suggesting that DFT cannot properly handle the physics of these compounds. 

Here, we show that DFT calculations correctly reproduce the experimental optical spectra of nickelates, properly accounting for the first peak and its shoulder as well as the second peak. Doing so, we further clarify that the second peak was erroneously assigned to the Peierls pseudo-gap. We highlight that the second peak is related to $t_{2g}$-$e_g$ transitions, missed in previous analysis based on a simple low-energy model including $e_g$ levels only.

\section{\label{sec:level2}Methods}

Our calculations are performed within the  DFT\cite{Hohenberg1964,Kohn1965} formalism as implemented in the {\sc Abinit} software package\cite{Gonze2002,Gonze2005,Gonze2009,Gonze2020}. We work within the Projected Augmented Wave (PAW) method \cite{Blochl1994}, relying on the atomic data from the JTH table\cite{Jollet2014}. The following valence and semi-core electrons are explicitly included in the calculations~: Y $5s^25p^65d^16s^2$, Nd $5s^25p^65d^16s^2$, Ni $3s^23p^63d^94s^1$ and O $2s^22p^4$. We use the GGA-PBE-sol\cite{Perdew2008} exchange-correlation functional. In order to better describe the electronic correlations of the localized Ni $3d$ electrons, a mean field Hubbard correction is included in the Lichtenstein formalism\cite{Lichtenstein1991} as implemented in {\sc Abinit} \cite{amadon2008}. We adopt the Hubbard correction $U=1.5$ eV which was previously shown  to provide accurate description together of the structural, magnetic and electronic properties of YNi$O_3$ \cite{Mercy2017}. The same value is then used for NdNiO$_3$, with the assumption that it is only slightly dependent of the tolerance factor\cite{Hampel2019}. Calculations are moreover performed within the collinear spin approximation.

We adopt a plane-wave basis set with an energy cut-off of 24 Ha and a grid of special $k$-points equivalent to $12\times12\times12$ in the 5-atoms unit cell. This guarantees a level of convergence of 1 meV/f.u. on the total energy. The relaxations are converged up to a maximum force of $5.10^{-5}$ Ha/bohr and a maximum stress of $5.10^{-7}$ Ha/bohr$^3$.
In order to correctly describe the insulating and metallic phases, we use a Fermi-Dirac occupation of the electronic states, with a smearing temperature of $300$ K.

The real part of the optical conductivity is computed using the Kubo-Greenwood formalism\cite{Mazevet2010}. Using {\sc Agate} [ref], the origin of the peaks is then related to specific electronic transitions that are analysed from density of states projected on specific atomic orbitals. From that, an interpretation is provided in terms of $e_g$/$t_{2g}$ transitions. Although conventional and appealing, this interpretation has to be taken with care and considered as only qualitative. Indeed, assessing the role of individual atomic orbital in the optical spectra is complicated by the amount of transitions (various $k$-points and energy bands) contributing to the optical spectrum at a given energy. Furthermore, the rotation and deformation of the oxygen cages make only approximate the decomposition into $e_g$/$t_{2g}$ states. Finally, we have to keep also in mind the strong hybridisation between Ni 3d and O 2p states. The most hybridized $t_{2g}$ and $e_g$ levels are marked by a $^\star$ in what follows.

\section{\label{sec:level3}Results and Discussion}

\subsection{Ground state properties}

Because of the strong interplay between structural, magnetic and electronic degrees of freedom in rare-earth nickelates, we cannot address their electronic and optical properties without evoking at first their structural and magnetic properties.

\begin{table}
\begin{tabular}{l>{\centering}m{1.4cm} >{\centering}m{1.4cm} >{\centering}m{1.4cm} c}
      \hline
       \hline
       Modes& \multicolumn{2}{c}{YNiO$_3$}   & \multicolumn{2}{c}{ NdNiO$_3$} \\
       & DFT & EXP & DFT & EXP \\
            \hline
     \textbf{$R_2^-$}  & 0.124 &0.127 & 0.091 &  0.072  \\
     \textbf{$R_5^-$} & 1.494 &1.492 & 1.040 &   1.146  \\
     \textbf{$M_2^+$}  & 1.132 &1.153  & 0.532 &  0.807 \\
    \hline
    \hline
    \end{tabular}
\caption{Comparison of the theoretical (DFT) and experimental (EXP) amplitudes (\AA) of the dominant lattice distortions with respect to the cubic reference in the $P2_1/n$ AFM-E' phase of YNiO$_3$ ($t=0.92$) and NdNiO$_3$ ($t=0.96$), as quantified with {\sc Amplimodes} \cite{Perez-Mato2010,Orobengoa2009}}.
  \label{modefigures}
\end{table}

The high-temperature prototypical reference phase of RNiO$_3$ nickelates is a priori the $Pm\bar{3}m$ cubic perovskite structure but it should only appear at very high temperatures and has never been observed experimentally. The observed high-temperature  metallic phase is of $Pbnm$ symmetry. It appears as a small distortion of the cubic reference involving mainly rotations ($M_3^+$ mode) and tilts ($R_5^-$ mode) of the NiO$_6$ octahedra  (see Table \ref{modefigures} for NdNiO$_3$ and YNiO$_3$). At the MIT, the system evolves from $Pbnm$ to  $P2_1/n$ symmetry. This $P2_1/n$ phase differs from the $Pbnm$ phase by the appearance of the breathing distortion of the oxygen cages ($R_2^-$ mode) yielding the appearance of large ($Ni_L$) and small ($Ni_S$) NiO$_6$ cages distributed according to a rocksalt pattern.

Here, all the  calculations are done in the $E'$ antiferromagnetic (AFM-E') ground state configuration corresponding to a ($\uparrow$, 0, $\downarrow$, 0) spin arrangement with a propagator vector of ($\frac{1}{2}$,$0$,$\frac{1}{2}$) in the $Pbnm$ setting. 
As previously discussed \cite{Mercy2017}, DFT calculations predict the correct magnetic and structural ground states. Moreover,  they provide very accurate description of the structural distortions (see Table \ref{modefigures}) and the obtained values for the magnetic exchange parameters give a Néel temperature in good agreement with experimental data\cite{Mercy2017}.

In the cubic reference phase, according to the octahedral environment of the nickel atoms, the crystal field already splits the Ni 3d levels into three lower-energy $t_{2g}$ and two higher-energy $e_g$ states. In the $Pbnm$ and $P2_1/n$ phases, oxygen rotations and tilts and related cell-shape deformations further increase the splitting between $t_{2g}$ and $e_g$ states. The $t_{2g} - e_g$  splitting being linked both to the Ni-O distance and the amplitude of the structural distortions, it increases from large to small rare-earth atom compounds.

\begin{figure}
 \includegraphics[width=1.05\linewidth]{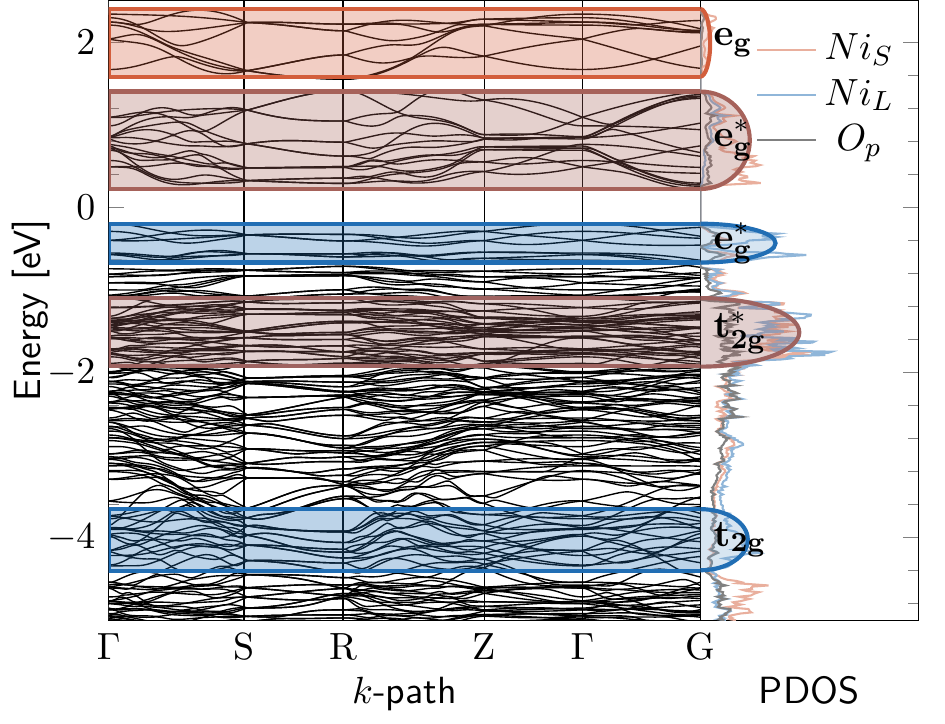}
 \caption{Electronic band structure of the insulating $P2_1/n$ AFM-E' phase of YNiO$_3$ as obtained from DFT calculations and related projected density of states (pDOS). For oxygen, the projection is illustrative and made on a single atom. Interpretation in terms of $e_g$ and $t_{2g}$ states is highlighted thanks to colored ellipse. The color is defined from the dominant large (blue) or small (red) nickel contribution.
  }\label{structband}
\end{figure}

We focus here on the optical spectra of the insulating $P2_1/n$ AFM-E' phase. The electronic band structure in this phase and related projected density of states are depicted in Figure \ref{structband} for YNiO$_3$. Although some minor differences can appear in the distributions of the highest $e_g$ states, this figure is representative of DFT calculations. In this phase, the valence states close to the Fermi level are dominantly $e_g$ states at Ni$_L$ sites. Moreover, in such DFT calculations, the lowest conduction states are dominantly $e_g$ states of the Ni$_S$ site while, contrary to another study, empty $e_g$ states  at Ni$_L$ are at higher energy. The gap does therefore not appear as a Mott gap between $e_g$ states at Ni$_L$ but rather as a Peierls gap between $e_g$ states from Ni$_L$ to Ni$_S$, directly proportional to the amplitude of the breathing distortion. Since the latter is triggered by rotation and tilt motions, the gap increases as the tolerance factor decreases. We further notice that activation of the breathing distortion also opens a pseudo-gap in the upper $e_g$ conduction bands. 

\subsection{\label{subsec:level1}Optical Spectra}

Since the electronic band structure of nickelates is sensitive to the tolerance factor,  we compare now optical spectra for representative compounds with small and large rare-earth cations. We consider at first NdNiO$_3$ ($t=0.96$) and compare our results to experimental data from Ruppen et al.\cite{Ruppen2015}. Then, we study the case of YNiO$_3$ ($t=0.92$) and compare the spectra of both compounds, relating changes to the evolution of the band structure.

\subsubsection{\label{subsubsect:level1} NdNiO$_3$}

\begin{figure}
 \includegraphics[width=0.95\linewidth]{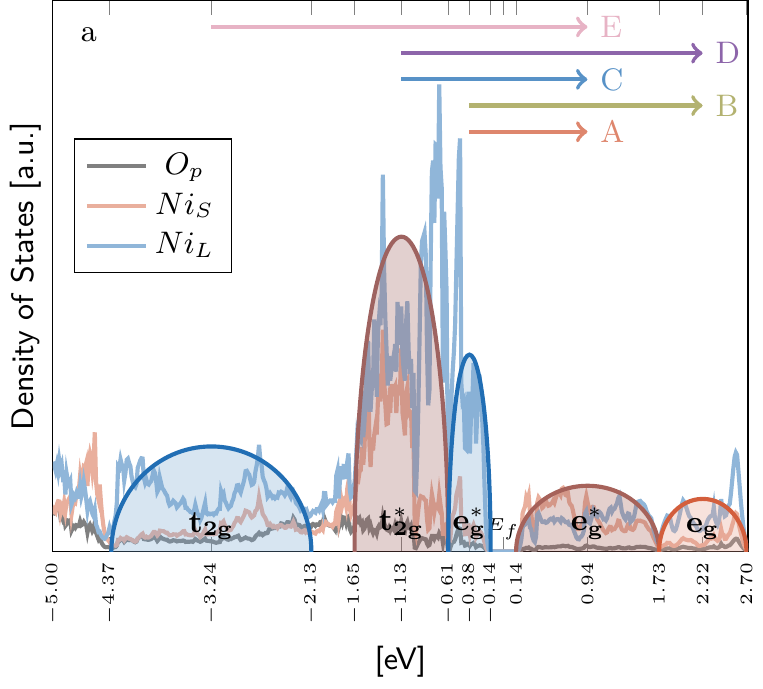}
  \includegraphics[width=0.95\linewidth]{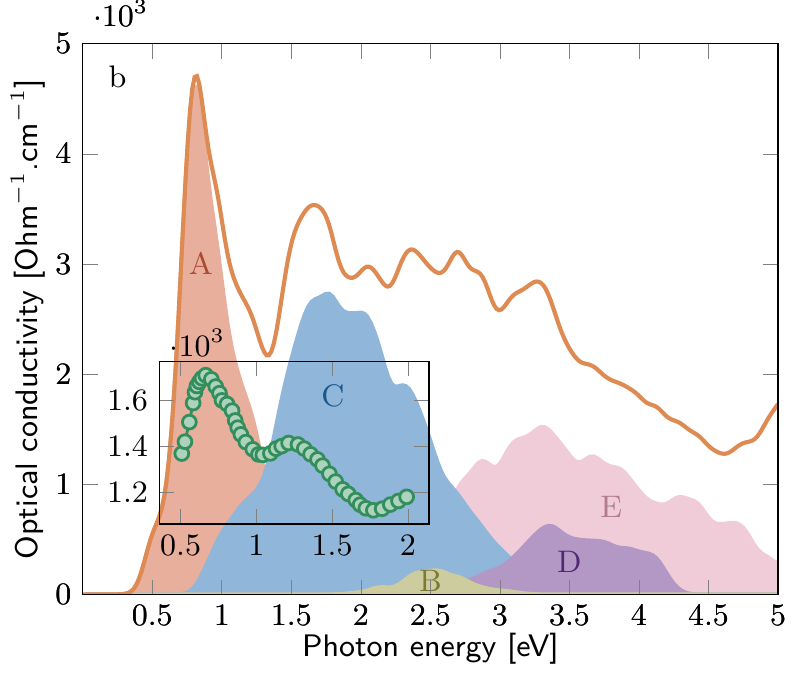}
    \caption{ (a) Projected DOS of NdNiO$_3$ in its $P2_1/n$ AFM-E' ground state, highlighting the main electronic transitions contributing to the optical spectrum. 
    Only one oxygen contribution is shown.
    (b) Optical Spectrum of NdNiO$_3$ in its $P2_1/n$ AFM-E' ground state and related individual contributions, as labelled in (a). Inset: Experimental optical spectrum at 100K\cite{Ruppen2015}.
  }\label{NdDOS}
\end{figure}

The projected electronic density of states around the Fermi level and the optical spectrum of NdNiO$_3$ are reported in  Fig. \ref{NdDOS}. We identify a first main peak (peak A in red) in the optical spectrum at an energy of about 0.8 eV, consistently with the value of the energy gap. This peak is associated to transitions from the occupied $e_g^*$ states of $Ni_L$ to the unoccupied $e_g^*$ states of $Ni_S$ (across the Peierls gap) as shown in the projected band structure on Fig. \ref{NdDOS}\textit{a}. Then, a second peak (peak C in blue) appears at an energy of about 1.6 eV and is related to transitions from $t_{2g}^*$ states at $Ni_S$ to the lowest unoccupied $e_g^*$ states of $Ni_S$. 

This assignment of the first and second peaks is different from the intepretation of Ruppen {\it et al.} \cite{Ruppen2015} who were relating the first peak to transitions between $e_g$ states at $Ni_L$ (Mott gap) and the second peak to transitions between $e_g$ states at $Ni_L$ and $Ni_S$ (Peierls pseudogap). 

It is worth mentioning that our DFT calculations also reproduce a pseudogap in the unoccupied $e_g$ states and we identify a peak (B in green) related to transitions from $e_g^*$ states at $Ni_L$ to $e_g$ states above the pseudo-gap, in the same range of energy, but the intensity related to these transitions is much too small to explain alone the second peak.

Then, we finally identify additional peaks at higher energies, originating from two main kinds of transitions.  The peaks D (purple) and E (pink) in figure \ref{NdDOS} correspond to transitions from occupied $t_{2g}^*$ at $Ni_S$ to $e_g$ states above the pseudo-gap and from deeper $t_{2g}$ states at $Ni_L$ to $e_g^*$ just above the gap, respectively. 

Comparison with experimental data (inset in Fig. \ref{NdDOS}b) highlights that, contrary to what was sometimes suggested\cite{Peil2019}, DFT optical spectra are not incompatible with measurements. They not only reproduce the first and the second peaks but even the intriguing shoulder observed in the second part of the first peak. From our calculations, this shoulder is related, on the one hand, to transitions from occupied $e_g^*$ states to the upper part of empty $e_g^*$ states and, on the other hand, to transitions from the states between the labelled $e_g^*$ and $t_{2g}^*$ bands (more related to the $t_{2g}^*$ states) and the empty $e_g^*$ bands.

It should moreover be noticed that, as previously discussed, the fixed value of U used in the present study was not optimized for NdNiO$_3$ but simply transferred from YNiO$_3$. As highlighted in Table 1, this yields a small overestimate of the breathing distortion. Slightly reducing U, would reduce the breathing distortion and the bandgap and would yield slightly more spread peaks,  still improving the quantitative agreement with experiment.

Although it was not necessarily guarantee nor obvious, this illustrates that the DFT band structure of rare-earth nickelates is compatible with optical measurements on these compounds. It also highlights the importance of considering explicitly $t_{2g}$ states while interpreting these data, which was missing in some previous models restricted to $e_g$ states.

\subsubsection{\label{subsubsect:level2} YNiO$_3$ and comparison}

\begin{figure}
 \includegraphics[width=0.95\linewidth]{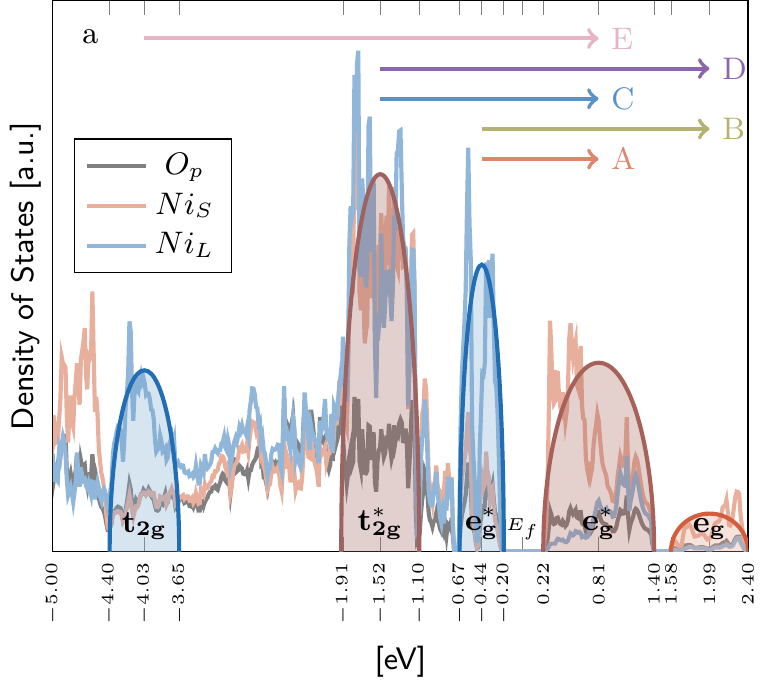}
  \includegraphics[width=0.95\linewidth]{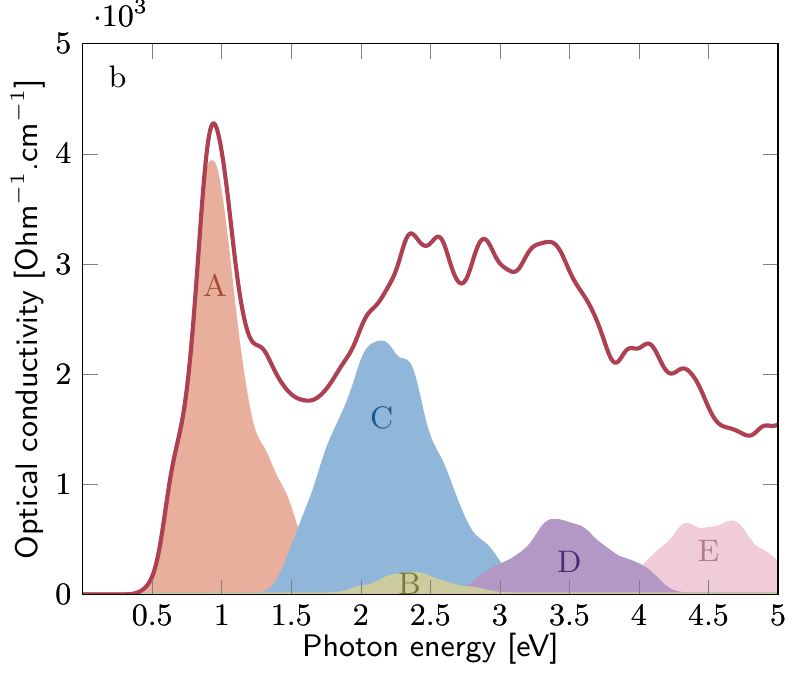}
  \caption{ (a) Projected DOS of YNiO$_3$ in its $P2_1/n$ AFM-E' ground state, highlighting the main electronic transitions contributing to the optical spectrum. 
 Only one oxygen contribution is shown.
    (b) Optical spectrum of YNiO$_3$ in its $P2_1/n$ AFM-E' ground state and related contributions as labelled in (a).
  }\label{YDOS}
\end{figure}

The projected electronic density of states around the Fermi level and the optical spectrum of YNiO$_3$ are reported in Fig. \ref{YDOS}. The shape and interpretation of the peaks are similar to NdNiO$_3$. Now, the location of the peaks is slightly modified according to the evolution of the electronic band structure. 

\begin{figure}
  \includegraphics[width=0.95\linewidth]{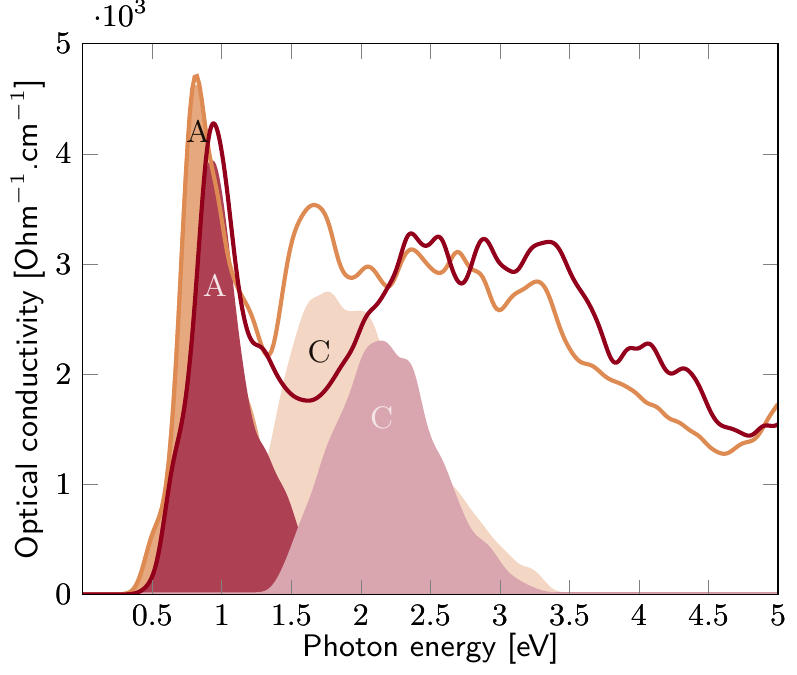}
    \caption{ Comparison of the optical spectra of NdNiO$_3$ (orange) and YNiO$_3$ (red) in their $P2_1/n$ AFM-E' ground state.
  }\label{spectre}
\end{figure}

Since the breathing distortion (Table \ref{modefigures}) and the bandgap are larger for small rare-earth atoms, the first peak appears at a slightly larger energy for YNiO$_3$ than NdNiO$_3$ (see Fig. \ref{spectre} for the comparison). This is in line with the experimental evolution of the optical spectrum from NdNiO$_3$ to SmNiO$_3$ \cite{Ruppen2015}.

Then, the evolution of the second peak is even more drastic. Moving from NdNiO$_3$ to YNiO$_3$, the Ni--O distances are reduced and the oxygen rotations are increased yielding a larger $t_{2g}-e_g$ splitting and moving the C peak at larger energy. 

Finally, as can be seen in Fig. \ref{spectre}, due to various other additional contributions, the modification of the full spectrum from NdNiO$_3$ to YNiO$_3$ is even more dramatic than what would have been be anticipated from the shift of the A and C peaks only.

\section{\label{sec:level4}Conclusion}

In this paper, we have reported the optical spectra of rare-earth nickelates, deduced from the DFT electronic band structure reproducing the bandgap as a Peierls gap.  Contrary to what was recently suggested \cite{Peil2019}, we show that the optical spectrum of NdNiO$_3$ is in fair agreement with available experimental data and reassigned the peaks highlighting the role of $t_{2g}$-$e_g$ transitions neglected from previous models. We also highlight significant change of the position of the second peak from large to small rare-earth nickelates. 

\section{\label{sec:level5}Acknowledgement}
\input acknowledgement.tex
\bibliography{biblio.bib}

\end{document}

%% file: author_list.tex
\author{Jordan Bieder}  \affiliation{Theoretical Materials Physics, Q-Mat, CESAM, University of Li\`ege, B-4000 Li\`ege, Belgium} 
\author{Alain Mercy} \affiliation{Theoretical Materials Physics, Q-Mat, CESAM, University of Li\`ege, B-4000 Li\`ege, Belgium}
\author{Wen-Yi Tong} \affiliation{Theoretical Materials Physics, Q-Mat, CESAM, University of Li\`ege, B-4000 Li\`ege, Belgium}
\author{Philippe Ghosez} \affiliation{Theoretical Materials Physics, Q-Mat, CESAM, University of Li\`ege, B-4000 Li\`ege, Belgium}

%
%
%
\vskip 0.25cm

%% file: acknowledgement.tex
%
We thank Vanina Recoules for helpful discussions on spectroscopy. This work was supported by the ARC project AIMED from University of Li\`ege and the PDR project PROMOSPAN from F.R.S.-FNRS Belgium. Computational resources were provided by the Consortium des Equipements de Calcul Intensif (CECI), funded by the F.R.S.-FNRS under the Grant No. 2.5020.11 and the Tier-1 supercomputer of the F\'ed\'eration Wallonie-Bruxelles funded by the Walloon Region under the Grant No. 1117545.

A.M. and J.B. contributed equally to this work.

%% file: main.bbl
\begin{thebibliography}{41}
\expandafter\ifx\csname natexlab\endcsname\relax\def\natexlab#1{#1}\fi
\expandafter\ifx\csname bibnamefont\endcsname\relax
  \def\bibnamefont#1{#1}\fi
\expandafter\ifx\csname bibfnamefont\endcsname\relax
  \def\bibfnamefont#1{#1}\fi
\expandafter\ifx\csname citenamefont\endcsname\relax
  \def\citenamefont#1{#1}\fi
\expandafter\ifx\csname url\endcsname\relax
  \def\url#1{\texttt{#1}}\fi
\expandafter\ifx\csname urlprefix\endcsname\relax\def\urlprefix{URL }\fi
\providecommand{\bibinfo}[2]{#2}
\providecommand{\eprint}[2][]{\url{#2}}

\bibitem[{\citenamefont{Khomskii}(2014)}]{Khomskii2014}
\bibinfo{author}{\bibfnamefont{D.~I.} \bibnamefont{Khomskii}},
  \emph{\bibinfo{title}{Transition Metal Coumpounds}}
  (\bibinfo{publisher}{Cambridge University Press}, \bibinfo{year}{2014}), ISBN
  \bibinfo{isbn}{9781107020177},
  \urlprefix\url{https://doi.org/10.1107/S2052520616003097}.

\bibitem[{\citenamefont{Imada et~al.}(1998)\citenamefont{Imada, Fujimori, and
  Tokura}}]{Imada1998}
\bibinfo{author}{\bibfnamefont{M.}~\bibnamefont{Imada}},
  \bibinfo{author}{\bibfnamefont{A.}~\bibnamefont{Fujimori}}, \bibnamefont{and}
  \bibinfo{author}{\bibfnamefont{Y.}~\bibnamefont{Tokura}},
  \bibinfo{journal}{Rev. Mod. Phys.} \textbf{\bibinfo{volume}{70}},
  \bibinfo{pages}{1039} (\bibinfo{year}{1998}),
  \urlprefix\url{https://link.aps.org/doi/10.1103/RevModPhys.70.1039}.

\bibitem[{\citenamefont{Zubko et~al.}(2011)\citenamefont{Zubko, Gariglio,
  Gabay, Ghosez, and Triscone}}]{Zubko2011}
\bibinfo{author}{\bibfnamefont{P.}~\bibnamefont{Zubko}},
  \bibinfo{author}{\bibfnamefont{S.}~\bibnamefont{Gariglio}},
  \bibinfo{author}{\bibfnamefont{M.}~\bibnamefont{Gabay}},
  \bibinfo{author}{\bibfnamefont{P.}~\bibnamefont{Ghosez}}, \bibnamefont{and}
  \bibinfo{author}{\bibfnamefont{J.-M.} \bibnamefont{Triscone}},
  \bibinfo{journal}{Annu. Rev. Conden. Ma. P.} \textbf{\bibinfo{volume}{2}},
  \bibinfo{pages}{141} (\bibinfo{year}{2011}),
  \urlprefix\url{https://doi.org/10.1146/annurev-conmatphys-062910-140445}.

\bibitem[{\citenamefont{Yuan et~al.}(2018)\citenamefont{Yuan, Lu, Stone, Wang,
  Brooks, Schlom, Sinnott, Zhou, and Gopalan}}]{Yuan2018}
\bibinfo{author}{\bibfnamefont{Y.}~\bibnamefont{Yuan}},
  \bibinfo{author}{\bibfnamefont{Y.}~\bibnamefont{Lu}},
  \bibinfo{author}{\bibfnamefont{G.}~\bibnamefont{Stone}},
  \bibinfo{author}{\bibfnamefont{K.}~\bibnamefont{Wang}},
  \bibinfo{author}{\bibfnamefont{C.~M.} \bibnamefont{Brooks}},
  \bibinfo{author}{\bibfnamefont{D.~G.} \bibnamefont{Schlom}},
  \bibinfo{author}{\bibfnamefont{S.~B.} \bibnamefont{Sinnott}},
  \bibinfo{author}{\bibfnamefont{H.}~\bibnamefont{Zhou}}, \bibnamefont{and}
  \bibinfo{author}{\bibfnamefont{V.}~\bibnamefont{Gopalan}},
  \bibinfo{journal}{Nature Communications} \textbf{\bibinfo{volume}{9}},
  \bibinfo{pages}{5220} (\bibinfo{year}{2018}),
  \urlprefix\url{https://doi.org/10.1038/s41467-018-07665-1}.

\bibitem[{\citenamefont{Grisolia et~al.}(2016)\citenamefont{Grisolia, Varignon,
  Sanchez-Santolino, Arora, Valencia, Varela, Abrudan, Weschke, Schierle, Rault
  et~al.}}]{Grisolia2016}
\bibinfo{author}{\bibfnamefont{M.~N.} \bibnamefont{Grisolia}},
  \bibinfo{author}{\bibfnamefont{J.}~\bibnamefont{Varignon}},
  \bibinfo{author}{\bibfnamefont{G.}~\bibnamefont{Sanchez-Santolino}},
  \bibinfo{author}{\bibfnamefont{A.}~\bibnamefont{Arora}},
  \bibinfo{author}{\bibfnamefont{S.}~\bibnamefont{Valencia}},
  \bibinfo{author}{\bibfnamefont{M.}~\bibnamefont{Varela}},
  \bibinfo{author}{\bibfnamefont{R.}~\bibnamefont{Abrudan}},
  \bibinfo{author}{\bibfnamefont{E.}~\bibnamefont{Weschke}},
  \bibinfo{author}{\bibfnamefont{E.}~\bibnamefont{Schierle}},
  \bibinfo{author}{\bibfnamefont{J.~E.} \bibnamefont{Rault}},
  \bibnamefont{et~al.}, \bibinfo{journal}{Nature Physics}
  \textbf{\bibinfo{volume}{12}}, \bibinfo{pages}{484} (\bibinfo{year}{2016}),
  ISSN \bibinfo{issn}{1745-2473},
  \urlprefix\url{http://www.nature.com/nphys/journal/v12/n5/full/nphys3627.html}.

\bibitem[{\citenamefont{Hwang et~al.}(2012)\citenamefont{Hwang, Iwasa,
  Kawasaki, Keimer, Nagaosa, and Tokura}}]{Hwang2012}
\bibinfo{author}{\bibfnamefont{H.~Y.} \bibnamefont{Hwang}},
  \bibinfo{author}{\bibfnamefont{Y.}~\bibnamefont{Iwasa}},
  \bibinfo{author}{\bibfnamefont{M.}~\bibnamefont{Kawasaki}},
  \bibinfo{author}{\bibfnamefont{B.}~\bibnamefont{Keimer}},
  \bibinfo{author}{\bibfnamefont{N.}~\bibnamefont{Nagaosa}}, \bibnamefont{and}
  \bibinfo{author}{\bibfnamefont{Y.}~\bibnamefont{Tokura}},
  \bibinfo{journal}{Nature Materials} \textbf{\bibinfo{volume}{11}},
  \bibinfo{pages}{130} (\bibinfo{year}{2012}),
  \urlprefix\url{https://doi.org/10.1038/nmat3223}.

\bibitem[{\citenamefont{Scherwitzl et~al.}(2011)\citenamefont{Scherwitzl,
  Gariglio, Gabay, Zubko, Gibert, and Triscone}}]{Scherwitzl2011}
\bibinfo{author}{\bibfnamefont{R.}~\bibnamefont{Scherwitzl}},
  \bibinfo{author}{\bibfnamefont{S.}~\bibnamefont{Gariglio}},
  \bibinfo{author}{\bibfnamefont{M.}~\bibnamefont{Gabay}},
  \bibinfo{author}{\bibfnamefont{P.}~\bibnamefont{Zubko}},
  \bibinfo{author}{\bibfnamefont{M.}~\bibnamefont{Gibert}}, \bibnamefont{and}
  \bibinfo{author}{\bibfnamefont{J.-M.} \bibnamefont{Triscone}},
  \bibinfo{journal}{Phys. Rev. Lett.} \textbf{\bibinfo{volume}{106}},
  \bibinfo{pages}{246403} (\bibinfo{year}{2011}),
  \urlprefix\url{https://link.aps.org/doi/10.1103/PhysRevLett.106.246403}.

\bibitem[{\citenamefont{Torrance et~al.}(1992)\citenamefont{Torrance, Lacorre,
  Nazzal, Ansaldo, and Niedermayer}}]{Torrance1992}
\bibinfo{author}{\bibfnamefont{J.~B.} \bibnamefont{Torrance}},
  \bibinfo{author}{\bibfnamefont{P.}~\bibnamefont{Lacorre}},
  \bibinfo{author}{\bibfnamefont{A.~I.} \bibnamefont{Nazzal}},
  \bibinfo{author}{\bibfnamefont{E.~J.} \bibnamefont{Ansaldo}},
  \bibnamefont{and}
  \bibinfo{author}{\bibfnamefont{C.}~\bibnamefont{Niedermayer}},
  \bibinfo{journal}{Physical Review B} \textbf{\bibinfo{volume}{45}}
  (\bibinfo{year}{1992}),
  \urlprefix\url{http://link.aps.org/doi/10.1103/PhysRevB.45.8209}.

\bibitem[{\citenamefont{Medarde}(1997)}]{Medarde1997}
\bibinfo{author}{\bibfnamefont{M.~L.} \bibnamefont{Medarde}},
  \bibinfo{journal}{Journal of Physics: Condensed Matter}
  \textbf{\bibinfo{volume}{9}}, \bibinfo{pages}{1679} (\bibinfo{year}{1997}),
  ISSN \bibinfo{issn}{0953-8984},
  \urlprefix\url{http://stacks.iop.org/0953-8984/9/i=8/a=003}.

\bibitem[{\citenamefont{Catalan}(2008)}]{Catalan2008}
\bibinfo{author}{\bibfnamefont{G.}~\bibnamefont{Catalan}},
  \bibinfo{journal}{Phase Transitions} \textbf{\bibinfo{volume}{81}},
  \bibinfo{pages}{729} (\bibinfo{year}{2008}), ISSN \bibinfo{issn}{0141-1594},
  \urlprefix\url{http://dx.doi.org/10.1080/01411590801992463}.

\bibitem[{\citenamefont{Goldschmidt}(1926)}]{Goldschmidt1926}
\bibinfo{author}{\bibfnamefont{V.~M.} \bibnamefont{Goldschmidt}},
  \bibinfo{journal}{Naturwissenschaften} \textbf{\bibinfo{volume}{14}},
  \bibinfo{pages}{477} (\bibinfo{year}{1926}).

\bibitem[{\citenamefont{Barman et~al.}(1994)\citenamefont{Barman, Chainani, and
  Sarma}}]{Barman1994}
\bibinfo{author}{\bibfnamefont{S.~R.} \bibnamefont{Barman}},
  \bibinfo{author}{\bibfnamefont{A.}~\bibnamefont{Chainani}}, \bibnamefont{and}
  \bibinfo{author}{\bibfnamefont{D.~D.} \bibnamefont{Sarma}},
  \bibinfo{journal}{Phys. Rev. B} \textbf{\bibinfo{volume}{49}},
  \bibinfo{pages}{8475} (\bibinfo{year}{1994}),
  \urlprefix\url{https://link.aps.org/doi/10.1103/PhysRevB.49.8475}.

\bibitem[{\citenamefont{Garcia-Munoz et~al.}(1992)\citenamefont{Garcia-Munoz,
  Rodriguez-Carvajal, Lacorre, and Torrance}}]{Garcia-Munoz1992}
\bibinfo{author}{\bibfnamefont{J.~L.} \bibnamefont{Garcia-Munoz}},
  \bibinfo{author}{\bibfnamefont{J.}~\bibnamefont{Rodriguez-Carvajal}},
  \bibinfo{author}{\bibfnamefont{P.}~\bibnamefont{Lacorre}}, \bibnamefont{and}
  \bibinfo{author}{\bibfnamefont{J.~B.} \bibnamefont{Torrance}},
  \bibinfo{journal}{Phys. Rev. B} \textbf{\bibinfo{volume}{46}},
  \bibinfo{pages}{4414} (\bibinfo{year}{1992}),
  \urlprefix\url{https://link.aps.org/doi/10.1103/PhysRevB.46.4414}.

\bibitem[{\citenamefont{Catalano et~al.}(2018)\citenamefont{Catalano, Gibert,
  Fowlie, {\'{I}}{\~{n}}iguez, Triscone, and Kreisel}}]{Catalano2018}
\bibinfo{author}{\bibfnamefont{S.}~\bibnamefont{Catalano}},
  \bibinfo{author}{\bibfnamefont{M.}~\bibnamefont{Gibert}},
  \bibinfo{author}{\bibfnamefont{J.}~\bibnamefont{Fowlie}},
  \bibinfo{author}{\bibfnamefont{J.}~\bibnamefont{{\'{I}}{\~{n}}iguez}},
  \bibinfo{author}{\bibfnamefont{J.-M.} \bibnamefont{Triscone}},
  \bibnamefont{and} \bibinfo{author}{\bibfnamefont{J.}~\bibnamefont{Kreisel}},
  \bibinfo{journal}{Reports on Progress in Physics}
  \textbf{\bibinfo{volume}{81}}, \bibinfo{pages}{046501}
  (\bibinfo{year}{2018}),
  \urlprefix\url{https://doi.org/10.1088%2F1361-6633%2Faaa37a}.

\bibitem[{\citenamefont{Mizokawa et~al.}(2000)\citenamefont{Mizokawa, Khomskii,
  and Sawatzky}}]{Mizokawa2000}
\bibinfo{author}{\bibfnamefont{T.}~\bibnamefont{Mizokawa}},
  \bibinfo{author}{\bibfnamefont{D.~I.} \bibnamefont{Khomskii}},
  \bibnamefont{and} \bibinfo{author}{\bibfnamefont{G.~A.}
  \bibnamefont{Sawatzky}}, \bibinfo{journal}{Phys. Rev. B}
  \textbf{\bibinfo{volume}{61}}, \bibinfo{pages}{11263} (\bibinfo{year}{2000}),
  \urlprefix\url{https://link.aps.org/doi/10.1103/PhysRevB.61.11263}.

\bibitem[{\citenamefont{Mazin et~al.}(2007)\citenamefont{Mazin, Khomskii,
  Lengsdorf, Alonso, Marshall, Ibberson, Podlesnyak, Mart\'{\i}nez-Lope, and
  Abd-Elmeguid}}]{Mazin2007}
\bibinfo{author}{\bibfnamefont{I.~I.} \bibnamefont{Mazin}},
  \bibinfo{author}{\bibfnamefont{D.~I.} \bibnamefont{Khomskii}},
  \bibinfo{author}{\bibfnamefont{R.}~\bibnamefont{Lengsdorf}},
  \bibinfo{author}{\bibfnamefont{J.~A.} \bibnamefont{Alonso}},
  \bibinfo{author}{\bibfnamefont{W.~G.} \bibnamefont{Marshall}},
  \bibinfo{author}{\bibfnamefont{R.~M.} \bibnamefont{Ibberson}},
  \bibinfo{author}{\bibfnamefont{A.}~\bibnamefont{Podlesnyak}},
  \bibinfo{author}{\bibfnamefont{M.~J.} \bibnamefont{Mart\'{\i}nez-Lope}},
  \bibnamefont{and} \bibinfo{author}{\bibfnamefont{M.~M.}
  \bibnamefont{Abd-Elmeguid}}, \bibinfo{journal}{Phys. Rev. Lett.}
  \textbf{\bibinfo{volume}{98}}, \bibinfo{pages}{176406}
  (\bibinfo{year}{2007}),
  \urlprefix\url{https://link.aps.org/doi/10.1103/PhysRevLett.98.176406}.

\bibitem[{\citenamefont{Seth et~al.}(2017)\citenamefont{Seth, Peil, Pourovskii,
  Betzinger, Friedrich, Parcollet, Biermann, Aryasetiawan, and
  Georges}}]{Seth2017}
\bibinfo{author}{\bibfnamefont{P.}~\bibnamefont{Seth}},
  \bibinfo{author}{\bibfnamefont{O.~E.} \bibnamefont{Peil}},
  \bibinfo{author}{\bibfnamefont{L.}~\bibnamefont{Pourovskii}},
  \bibinfo{author}{\bibfnamefont{M.}~\bibnamefont{Betzinger}},
  \bibinfo{author}{\bibfnamefont{C.}~\bibnamefont{Friedrich}},
  \bibinfo{author}{\bibfnamefont{O.}~\bibnamefont{Parcollet}},
  \bibinfo{author}{\bibfnamefont{S.}~\bibnamefont{Biermann}},
  \bibinfo{author}{\bibfnamefont{F.}~\bibnamefont{Aryasetiawan}},
  \bibnamefont{and} \bibinfo{author}{\bibfnamefont{A.}~\bibnamefont{Georges}},
  \bibinfo{journal}{Phys. Rev. B} \textbf{\bibinfo{volume}{96}},
  \bibinfo{pages}{205139} (\bibinfo{year}{2017}),
  \urlprefix\url{https://link.aps.org/doi/10.1103/PhysRevB.96.205139}.

\bibitem[{\citenamefont{Park et~al.}(2012)\citenamefont{Park, Millis, and
  Marianetti}}]{Park2012}
\bibinfo{author}{\bibfnamefont{H.}~\bibnamefont{Park}},
  \bibinfo{author}{\bibfnamefont{A.~J.} \bibnamefont{Millis}},
  \bibnamefont{and} \bibinfo{author}{\bibfnamefont{C.~A.}
  \bibnamefont{Marianetti}}, \bibinfo{journal}{Physical Review Letters}
  \textbf{\bibinfo{volume}{109}}, \bibinfo{pages}{156402}
  (\bibinfo{year}{2012}),
  \urlprefix\url{http://link.aps.org/doi/10.1103/PhysRevLett.109.156402}.

\bibitem[{\citenamefont{Subedi et~al.}(2015)\citenamefont{Subedi, Peil, and
  Georges}}]{Subedi2015}
\bibinfo{author}{\bibfnamefont{A.}~\bibnamefont{Subedi}},
  \bibinfo{author}{\bibfnamefont{O.~E.} \bibnamefont{Peil}}, \bibnamefont{and}
  \bibinfo{author}{\bibfnamefont{A.}~\bibnamefont{Georges}},
  \bibinfo{journal}{Physical Review B} \textbf{\bibinfo{volume}{91}},
  \bibinfo{pages}{075128} (\bibinfo{year}{2015}),
  \urlprefix\url{http://link.aps.org/doi/10.1103/PhysRevB.91.075128}.

\bibitem[{\citenamefont{Mercy et~al.}(2017)\citenamefont{Mercy, Bieder,
  Iniguez, and Ghosez}}]{Mercy2017}
\bibinfo{author}{\bibfnamefont{A.}~\bibnamefont{Mercy}},
  \bibinfo{author}{\bibfnamefont{J.}~\bibnamefont{Bieder}},
  \bibinfo{author}{\bibfnamefont{J.}~\bibnamefont{Iniguez}}, \bibnamefont{and}
  \bibinfo{author}{\bibfnamefont{P.}~\bibnamefont{Ghosez}},
  \bibinfo{journal}{Nature Communications}  (\bibinfo{year}{2017}),
  \urlprefix\url{https://www.nature.com/articles/s41467-017-01811-x}.

\bibitem[{\citenamefont{Peil et~al.}(2019)\citenamefont{Peil, Hampel, Ederer,
  and Georges}}]{Peil2019}
\bibinfo{author}{\bibfnamefont{O.~E.} \bibnamefont{Peil}},
  \bibinfo{author}{\bibfnamefont{A.}~\bibnamefont{Hampel}},
  \bibinfo{author}{\bibfnamefont{C.}~\bibnamefont{Ederer}}, \bibnamefont{and}
  \bibinfo{author}{\bibfnamefont{A.}~\bibnamefont{Georges}},
  \bibinfo{journal}{Phys. Rev. B} \textbf{\bibinfo{volume}{99}},
  \bibinfo{pages}{245127} (\bibinfo{year}{2019}),
  \urlprefix\url{https://link.aps.org/doi/10.1103/PhysRevB.99.245127}.

\bibitem[{\citenamefont{Hampel et~al.}(2019)\citenamefont{Hampel, Liu,
  Franchini, and Ederer}}]{Hampel2019}
\bibinfo{author}{\bibfnamefont{A.}~\bibnamefont{Hampel}},
  \bibinfo{author}{\bibfnamefont{P.}~\bibnamefont{Liu}},
  \bibinfo{author}{\bibfnamefont{C.}~\bibnamefont{Franchini}},
  \bibnamefont{and} \bibinfo{author}{\bibfnamefont{C.}~\bibnamefont{Ederer}},
  \bibinfo{journal}{npj Quantum Materials} \textbf{\bibinfo{volume}{4}}
  (\bibinfo{year}{2019}),
  \urlprefix\url{https://doi.org/10.1038/s41535-019-0145-4}.

\bibitem[{\citenamefont{Ruppen et~al.}(2015)\citenamefont{Ruppen, Teyssier,
  Peil, Catalano, Gibert, Mravlje, Triscone, Georges, and van~der
  Marel}}]{Ruppen2015}
\bibinfo{author}{\bibfnamefont{J.}~\bibnamefont{Ruppen}},
  \bibinfo{author}{\bibfnamefont{J.}~\bibnamefont{Teyssier}},
  \bibinfo{author}{\bibfnamefont{O.~E.} \bibnamefont{Peil}},
  \bibinfo{author}{\bibfnamefont{S.}~\bibnamefont{Catalano}},
  \bibinfo{author}{\bibfnamefont{M.}~\bibnamefont{Gibert}},
  \bibinfo{author}{\bibfnamefont{J.}~\bibnamefont{Mravlje}},
  \bibinfo{author}{\bibfnamefont{J.-M.} \bibnamefont{Triscone}},
  \bibinfo{author}{\bibfnamefont{A.}~\bibnamefont{Georges}}, \bibnamefont{and}
  \bibinfo{author}{\bibfnamefont{D.}~\bibnamefont{van~der Marel}},
  \bibinfo{journal}{Physical Review B} \textbf{\bibinfo{volume}{92}},
  \bibinfo{pages}{155145} (\bibinfo{year}{2015}),
  \urlprefix\url{http://link.aps.org/doi/10.1103/PhysRevB.92.155145}.

\bibitem[{\citenamefont{Stewart et~al.}(2011)\citenamefont{Stewart, Liu,
  Kareev, Chakhalian, and Basov}}]{Stewart2011}
\bibinfo{author}{\bibfnamefont{M.~K.} \bibnamefont{Stewart}},
  \bibinfo{author}{\bibfnamefont{J.}~\bibnamefont{Liu}},
  \bibinfo{author}{\bibfnamefont{M.}~\bibnamefont{Kareev}},
  \bibinfo{author}{\bibfnamefont{J.}~\bibnamefont{Chakhalian}},
  \bibnamefont{and} \bibinfo{author}{\bibfnamefont{D.~N.} \bibnamefont{Basov}},
  \bibinfo{journal}{Phys. Rev. Lett.} \textbf{\bibinfo{volume}{107}},
  \bibinfo{pages}{176401} (\bibinfo{year}{2011}),
  \urlprefix\url{https://link.aps.org/doi/10.1103/PhysRevLett.107.176401}.

\bibitem[{\citenamefont{Torriss et~al.}(2017)\citenamefont{Torriss, Margot, and
  Chaker}}]{Toriss2017}
\bibinfo{author}{\bibfnamefont{B.}~\bibnamefont{Torriss}},
  \bibinfo{author}{\bibfnamefont{J.}~\bibnamefont{Margot}}, \bibnamefont{and}
  \bibinfo{author}{\bibfnamefont{M.}~\bibnamefont{Chaker}},
  \bibinfo{journal}{Scientific Reports} \textbf{\bibinfo{volume}{7}},
  \bibinfo{pages}{40915} (\bibinfo{year}{2017}),
  \urlprefix\url{https://doi.org/10.1038/srep40915}.

\bibitem[{\citenamefont{Georgescu et~al.}(2019)\citenamefont{Georgescu, Peil,
  Disa, Georges, and Millis}}]{Georgescu2019}
\bibinfo{author}{\bibfnamefont{A.~B.} \bibnamefont{Georgescu}},
  \bibinfo{author}{\bibfnamefont{O.~E.} \bibnamefont{Peil}},
  \bibinfo{author}{\bibfnamefont{A.~S.} \bibnamefont{Disa}},
  \bibinfo{author}{\bibfnamefont{A.}~\bibnamefont{Georges}}, \bibnamefont{and}
  \bibinfo{author}{\bibfnamefont{A.~J.} \bibnamefont{Millis}},
  \bibinfo{journal}{Proceedings of the National Academy of Sciences}
  \textbf{\bibinfo{volume}{116}}, \bibinfo{pages}{14434}
  (\bibinfo{year}{2019}), ISSN \bibinfo{issn}{0027-8424},
  \eprint{https://www.pnas.org/content/116/29/14434.full.pdf},
  \urlprefix\url{https://www.pnas.org/content/116/29/14434}.

\bibitem[{\citenamefont{Ruppen et~al.}(2017)\citenamefont{Ruppen, Teyssier,
  Ardizzone, Peil, Catalano, Gibert, Triscone, Georges, and van~der
  Marel}}]{Ruppen2017}
\bibinfo{author}{\bibfnamefont{J.}~\bibnamefont{Ruppen}},
  \bibinfo{author}{\bibfnamefont{J.}~\bibnamefont{Teyssier}},
  \bibinfo{author}{\bibfnamefont{I.}~\bibnamefont{Ardizzone}},
  \bibinfo{author}{\bibfnamefont{O.~E.} \bibnamefont{Peil}},
  \bibinfo{author}{\bibfnamefont{S.}~\bibnamefont{Catalano}},
  \bibinfo{author}{\bibfnamefont{M.}~\bibnamefont{Gibert}},
  \bibinfo{author}{\bibfnamefont{J.-M.} \bibnamefont{Triscone}},
  \bibinfo{author}{\bibfnamefont{A.}~\bibnamefont{Georges}}, \bibnamefont{and}
  \bibinfo{author}{\bibfnamefont{D.}~\bibnamefont{van~der Marel}},
  \bibinfo{journal}{Phys. Rev. B} \textbf{\bibinfo{volume}{96}},
  \bibinfo{pages}{045120} (\bibinfo{year}{2017}),
  \urlprefix\url{https://link.aps.org/doi/10.1103/PhysRevB.96.045120}.

\bibitem[{\citenamefont{Hohenberg and Kohn}(1964)}]{Hohenberg1964}
\bibinfo{author}{\bibfnamefont{P.}~\bibnamefont{Hohenberg}} \bibnamefont{and}
  \bibinfo{author}{\bibfnamefont{W.}~\bibnamefont{Kohn}},
  \bibinfo{journal}{Phys. Rev.} \textbf{\bibinfo{volume}{136}},
  \bibinfo{pages}{B864} (\bibinfo{year}{1964}),
  \urlprefix\url{http://link.aps.org/doi/10.1103/PhysRev.136.B864}.

\bibitem[{\citenamefont{Kohn and Sham}(1965)}]{Kohn1965}
\bibinfo{author}{\bibfnamefont{W.}~\bibnamefont{Kohn}} \bibnamefont{and}
  \bibinfo{author}{\bibfnamefont{L.~J.} \bibnamefont{Sham}},
  \bibinfo{journal}{Physical Review} \textbf{\bibinfo{volume}{140}},
  \bibinfo{pages}{A1133} (\bibinfo{year}{1965}),
  \urlprefix\url{http://link.aps.org/doi/10.1103/PhysRev.140.A1133}.

\bibitem[{\citenamefont{Gonze et~al.}(2002)\citenamefont{Gonze, Beuken,
  Caracas, Detraux, Fuchs, Rignanese, Sindic, Verstraete, Zerah, Jollet
  et~al.}}]{Gonze2002}
\bibinfo{author}{\bibfnamefont{X.}~\bibnamefont{Gonze}},
  \bibinfo{author}{\bibfnamefont{J.-M.} \bibnamefont{Beuken}},
  \bibinfo{author}{\bibfnamefont{R.}~\bibnamefont{Caracas}},
  \bibinfo{author}{\bibfnamefont{F.}~\bibnamefont{Detraux}},
  \bibinfo{author}{\bibfnamefont{M.}~\bibnamefont{Fuchs}},
  \bibinfo{author}{\bibfnamefont{G.-M.} \bibnamefont{Rignanese}},
  \bibinfo{author}{\bibfnamefont{L.}~\bibnamefont{Sindic}},
  \bibinfo{author}{\bibfnamefont{M.}~\bibnamefont{Verstraete}},
  \bibinfo{author}{\bibfnamefont{G.}~\bibnamefont{Zerah}},
  \bibinfo{author}{\bibfnamefont{F.}~\bibnamefont{Jollet}},
  \bibnamefont{et~al.}, \bibinfo{journal}{Computational Materials Science}
  \textbf{\bibinfo{volume}{25}}, \bibinfo{pages}{478 } (\bibinfo{year}{2002}),
  ISSN \bibinfo{issn}{0927-0256},
  \urlprefix\url{http://www.sciencedirect.com/science/article/pii/S0927025602003257}.

\bibitem[{\citenamefont{Gonze et~al.}(2005)\citenamefont{Gonze, Rignanese,
  Verstraete, Betiken, Pouillon, Caracas, Jollet, Torrent, Zerah, Mikami
  et~al.}}]{Gonze2005}
\bibinfo{author}{\bibfnamefont{X.}~\bibnamefont{Gonze}},
  \bibinfo{author}{\bibfnamefont{G.}~\bibnamefont{Rignanese}},
  \bibinfo{author}{\bibfnamefont{M.}~\bibnamefont{Verstraete}},
  \bibinfo{author}{\bibfnamefont{J.}~\bibnamefont{Betiken}},
  \bibinfo{author}{\bibfnamefont{Y.}~\bibnamefont{Pouillon}},
  \bibinfo{author}{\bibfnamefont{R.}~\bibnamefont{Caracas}},
  \bibinfo{author}{\bibfnamefont{F.}~\bibnamefont{Jollet}},
  \bibinfo{author}{\bibfnamefont{M.}~\bibnamefont{Torrent}},
  \bibinfo{author}{\bibfnamefont{G.}~\bibnamefont{Zerah}},
  \bibinfo{author}{\bibfnamefont{M.}~\bibnamefont{Mikami}},
  \bibnamefont{et~al.}, \bibinfo{journal}{Zeitschrift fur
  Kristallographie.(Special issue on Computational Crystallography.)}
  \textbf{\bibinfo{volume}{220}}, \bibinfo{pages}{558} (\bibinfo{year}{2005}).

\bibitem[{\citenamefont{Gonze et~al.}(2009)\citenamefont{Gonze, Amadon,
  Anglade, Beuken, Bottin, Boulanger, Bruneval, Caliste, Caracas, Côté
  et~al.}}]{Gonze2009}
\bibinfo{author}{\bibfnamefont{X.}~\bibnamefont{Gonze}},
  \bibinfo{author}{\bibfnamefont{B.}~\bibnamefont{Amadon}},
  \bibinfo{author}{\bibfnamefont{P.~M.} \bibnamefont{Anglade}},
  \bibinfo{author}{\bibfnamefont{J.~M.} \bibnamefont{Beuken}},
  \bibinfo{author}{\bibfnamefont{F.}~\bibnamefont{Bottin}},
  \bibinfo{author}{\bibfnamefont{P.}~\bibnamefont{Boulanger}},
  \bibinfo{author}{\bibfnamefont{F.}~\bibnamefont{Bruneval}},
  \bibinfo{author}{\bibfnamefont{D.}~\bibnamefont{Caliste}},
  \bibinfo{author}{\bibfnamefont{R.}~\bibnamefont{Caracas}},
  \bibinfo{author}{\bibfnamefont{M.}~\bibnamefont{Côté}},
  \bibnamefont{et~al.}, \bibinfo{journal}{Computer Physics Communications}
  \textbf{\bibinfo{volume}{180}}, \bibinfo{pages}{2582} (\bibinfo{year}{2009}),
  ISSN \bibinfo{issn}{0010-4655},
  \urlprefix\url{http://www.sciencedirect.com/science/article/pii/S0010465509002276}.

\bibitem[{\citenamefont{Gonze et~al.}(2020)\citenamefont{Gonze, Amadon,
  Antonius, Arnardi, Baguet, Beuken, Bieder, Bottin, Bouchet, Bousquet
  et~al.}}]{Gonze2020}
\bibinfo{author}{\bibfnamefont{X.}~\bibnamefont{Gonze}},
  \bibinfo{author}{\bibfnamefont{B.}~\bibnamefont{Amadon}},
  \bibinfo{author}{\bibfnamefont{G.}~\bibnamefont{Antonius}},
  \bibinfo{author}{\bibfnamefont{F.}~\bibnamefont{Arnardi}},
  \bibinfo{author}{\bibfnamefont{L.}~\bibnamefont{Baguet}},
  \bibinfo{author}{\bibfnamefont{J.-M.} \bibnamefont{Beuken}},
  \bibinfo{author}{\bibfnamefont{J.}~\bibnamefont{Bieder}},
  \bibinfo{author}{\bibfnamefont{F.}~\bibnamefont{Bottin}},
  \bibinfo{author}{\bibfnamefont{J.}~\bibnamefont{Bouchet}},
  \bibinfo{author}{\bibfnamefont{E.}~\bibnamefont{Bousquet}},
  \bibnamefont{et~al.}, \bibinfo{journal}{Computer Physics Communications}
  \textbf{\bibinfo{volume}{248}}, \bibinfo{pages}{107042}
  (\bibinfo{year}{2020}), ISSN \bibinfo{issn}{0010-4655},
  \urlprefix\url{http://www.sciencedirect.com/science/article/pii/S0010465519303741}.

\bibitem[{\citenamefont{Blöchl}(1994)}]{Blochl1994}
\bibinfo{author}{\bibfnamefont{P.~E.} \bibnamefont{Blöchl}},
  \bibinfo{journal}{Physical Review B} \textbf{\bibinfo{volume}{50}},
  \bibinfo{pages}{17953} (\bibinfo{year}{1994}),
  \urlprefix\url{http://link.aps.org/doi/10.1103/PhysRevB.50.17953}.

\bibitem[{\citenamefont{Jollet et~al.}(2014)\citenamefont{Jollet, Torrent, and
  Holzwarth}}]{Jollet2014}
\bibinfo{author}{\bibfnamefont{F.}~\bibnamefont{Jollet}},
  \bibinfo{author}{\bibfnamefont{M.}~\bibnamefont{Torrent}}, \bibnamefont{and}
  \bibinfo{author}{\bibfnamefont{N.}~\bibnamefont{Holzwarth}},
  \bibinfo{journal}{Computer Physics Communications}
  \textbf{\bibinfo{volume}{185}}, \bibinfo{pages}{1246} (\bibinfo{year}{2014}),
  ISSN \bibinfo{issn}{0010-4655},
  \urlprefix\url{http://www.sciencedirect.com/science/article/pii/S0010465513004359}.

\bibitem[{\citenamefont{Perdew et~al.}(2008)\citenamefont{Perdew, Ruzsinszky,
  Csonka, Vydrov, Scuseria, Constantin, Zhou, and Burke}}]{Perdew2008}
\bibinfo{author}{\bibfnamefont{J.~P.} \bibnamefont{Perdew}},
  \bibinfo{author}{\bibfnamefont{A.}~\bibnamefont{Ruzsinszky}},
  \bibinfo{author}{\bibfnamefont{G.~I.} \bibnamefont{Csonka}},
  \bibinfo{author}{\bibfnamefont{O.~A.} \bibnamefont{Vydrov}},
  \bibinfo{author}{\bibfnamefont{G.~E.} \bibnamefont{Scuseria}},
  \bibinfo{author}{\bibfnamefont{L.~A.} \bibnamefont{Constantin}},
  \bibinfo{author}{\bibfnamefont{X.}~\bibnamefont{Zhou}}, \bibnamefont{and}
  \bibinfo{author}{\bibfnamefont{K.}~\bibnamefont{Burke}},
  \bibinfo{journal}{Phys. Rev. Lett.} \textbf{\bibinfo{volume}{100}},
  \bibinfo{pages}{136406} (\bibinfo{year}{2008}),
  \urlprefix\url{http://link.aps.org/doi/10.1103/PhysRevLett.100.136406}.

\bibitem[{\citenamefont{Liechtenstein et~al.}(1991)\citenamefont{Liechtenstein,
  Mazin, Rodriguez, Jepsen, Andersen, and Methfessel}}]{Lichtenstein1991}
\bibinfo{author}{\bibfnamefont{A.~I.} \bibnamefont{Liechtenstein}},
  \bibinfo{author}{\bibfnamefont{I.~I.} \bibnamefont{Mazin}},
  \bibinfo{author}{\bibfnamefont{C.~O.} \bibnamefont{Rodriguez}},
  \bibinfo{author}{\bibfnamefont{O.}~\bibnamefont{Jepsen}},
  \bibinfo{author}{\bibfnamefont{O.~K.} \bibnamefont{Andersen}},
  \bibnamefont{and}
  \bibinfo{author}{\bibfnamefont{M.}~\bibnamefont{Methfessel}},
  \bibinfo{journal}{Phys. Rev. B} \textbf{\bibinfo{volume}{44}},
  \bibinfo{pages}{5388} (\bibinfo{year}{1991}),
  \urlprefix\url{https://link.aps.org/doi/10.1103/PhysRevB.44.5388}.

\bibitem[{\citenamefont{Amadon et~al.}(2008)\citenamefont{Amadon, Lechermann,
  Georges, Jollet, Wehling, and Lichtenstein}}]{amadon2008}
\bibinfo{author}{\bibfnamefont{B.}~\bibnamefont{Amadon}},
  \bibinfo{author}{\bibfnamefont{F.}~\bibnamefont{Lechermann}},
  \bibinfo{author}{\bibfnamefont{A.}~\bibnamefont{Georges}},
  \bibinfo{author}{\bibfnamefont{F.}~\bibnamefont{Jollet}},
  \bibinfo{author}{\bibfnamefont{T.~O.} \bibnamefont{Wehling}},
  \bibnamefont{and} \bibinfo{author}{\bibfnamefont{A.~I.}
  \bibnamefont{Lichtenstein}}, \bibinfo{journal}{Phys. Rev. B}
  \textbf{\bibinfo{volume}{77}}, \bibinfo{pages}{205112}
  (\bibinfo{year}{2008}),
  \urlprefix\url{https://link.aps.org/doi/10.1103/PhysRevB.77.205112}.

\bibitem[{\citenamefont{Mazevet et~al.}(2010)\citenamefont{Mazevet, Torrent,
  Recoules, and Jollet}}]{Mazevet2010}
\bibinfo{author}{\bibfnamefont{S.}~\bibnamefont{Mazevet}},
  \bibinfo{author}{\bibfnamefont{M.}~\bibnamefont{Torrent}},
  \bibinfo{author}{\bibfnamefont{V.}~\bibnamefont{Recoules}}, \bibnamefont{and}
  \bibinfo{author}{\bibfnamefont{F.}~\bibnamefont{Jollet}},
  \bibinfo{journal}{High Energy Density Physics} \textbf{\bibinfo{volume}{6}},
  \bibinfo{pages}{84 } (\bibinfo{year}{2010}), ISSN \bibinfo{issn}{1574-1818},
  \urlprefix\url{http://www.sciencedirect.com/science/article/pii/S1574181809000664}.

\bibitem[{\citenamefont{Perez-Mato et~al.}(2010)\citenamefont{Perez-Mato,
  Orobengoa, and Aroyo}}]{Perez-Mato2010}
\bibinfo{author}{\bibfnamefont{J.~M.} \bibnamefont{Perez-Mato}},
  \bibinfo{author}{\bibfnamefont{D.}~\bibnamefont{Orobengoa}},
  \bibnamefont{and} \bibinfo{author}{\bibfnamefont{M.~I.} \bibnamefont{Aroyo}},
  \bibinfo{journal}{Acta Crystallographica Section A}
  \textbf{\bibinfo{volume}{66}}, \bibinfo{pages}{558} (\bibinfo{year}{2010}),
  \urlprefix\url{https://doi.org/10.1107/S0108767310016247}.

\bibitem[{\citenamefont{Orobengoa et~al.}(2009)\citenamefont{Orobengoa,
  Capillas, Aroyo, and Perez-Mato}}]{Orobengoa2009}
\bibinfo{author}{\bibfnamefont{D.}~\bibnamefont{Orobengoa}},
  \bibinfo{author}{\bibfnamefont{C.}~\bibnamefont{Capillas}},
  \bibinfo{author}{\bibfnamefont{M.~I.} \bibnamefont{Aroyo}}, \bibnamefont{and}
  \bibinfo{author}{\bibfnamefont{J.~M.} \bibnamefont{Perez-Mato}},
  \bibinfo{journal}{Journal of Applied Crystallography}
  \textbf{\bibinfo{volume}{42}}, \bibinfo{pages}{820} (\bibinfo{year}{2009}),
  \urlprefix\url{https://doi.org/10.1107/S0021889809028064}.

\end{thebibliography}
